# Relative Dating and Classification of Minerals and Rocks Based on Statistical Calculations Related to Their Potential Energy Index

Mikhail M. Labushev
Department of Ore Deposits Geology and Prospecting, Institute of Mining, Geology and Geotechnology, Siberian Federal University, Krasnoyarsk, Russian Federation
Alexander N. Khokhlov
JSC "Geoeconomy", Krasnoyarsk, Russian Federation

## ABSTRACT

Index of proportionality of atomic weights of chemical elements is proposed for determining the relative age of minerals and rocks. Their chemical analysis results serve to be initial data for calculations. For rocks of different composition the index is considered to be classification value as well. Crystal lattice energy change in minerals and their associations can be measured by the index value change, thus contributing to the solution of important practical problems. There was determined the effect of more rapid increase of potential energy of limestone with relatively low lattice energy as compared with the others.

## INTRODUCTION

The history of relative dating of rocks is associated with the names of Nicolas Steno, 1669; James Hutton, 1795; Georges Cuvier and Alexandre Brongniart, 1811; William Smith, 1815; Charles Lyell, 1830 and Henry Clifton Sorby, 1858. To determine relative age of rocks the following principles of relative chronology [1, 2, 3, and 4] are used:

- superposition
- original horizontality
- lateral continuity
- cross-cutting relationships
- intrusive relationships
- inclusions of igneous rocks
- inclusions and components
- faunal succession
- included fragments

We note that the use of these principles assumes the measurement at nominal or ordinal scales of measurement. New developments in the relative dating field seem to have a large scientific and practical significance, since the well-known principles have different conditions of application and therefore can complement each other and be used to verify the findings. New principles, using ratio or interval scales are particularly important.

We propose the definition of the relative age of rocks, based on the calculation of the index of proportionality of the atomic weights of chemical elements of minerals and rocks in the interval scale of measurement.

## METHOD

There are three approaches to the energy stability of crystals considering:

- ionic lattice energy (energy required to break crystal into free ions)
- separation into individual atoms (energy required to break crystal into free neutral atoms)
- bonding frames and elektrids [5]

The third approach suggests that a crystal contains atomic frames consisting of nuclei and internal electrons, being not involved in chemical bonding, and elektrid composed of bonding or shared valence electrons as well as nonbonding or unshared ones. In this case we can approximately calculate the energy of any chemical bond in crystals regardless of its type.

As the result of additions and clarifications of the ion lattice classical formula [6, 7], there was derived a formula of adhesion energy of frames and elektrids (MJ/mol) for the general case of crystalline compounds [5]:

$$W = -\frac{1.389\alpha^{R} q^{0} q^{e} \tau p}{d}$$

where: 1.389 – constant for the expression of the adhesion energy of frames and elektrids in MJ/mol; $\alpha^R$ – the reduced Madelung constant; $q^0$ –total effective nuclear charge acting on the valence electrons of atom; $q^e$ – charge of elektrid; $\tau$ – repulsion coefficient equal to 0.1-0.7; $p$ - coefficient characterizing bond covalence, equal to $(1-f_i^2)^{1/2}$, where $f_i$ – bond ionicity; $d$ - metal - nonmetal internuclear distance in crystal. For complex crystalline compounds there should be used averaged parameters $q^0$, $q^e$, $f_i$, $d$.

The formula shows the inversely proportional dependence of the potential energy of crystals and their associations on metal - nonmetal internuclear distance. It is of importance that instead of the changeable the Madelung constant the formula contains the less variable the reduced Madelung constant. These calculations are of practical interest for the study of minerals and rocks. The change of crystal lattice energy in mineral associations can be probably measured by the averaged internuclear distances change, thus contributing to the solution of important practical problems.

For this purpose it is proposed to calculate $I_{av}$ index [8] as well. It was established experimentally that the values of the index for minerals and rocks are also inversely proportional to the energy of crystal lattice while being directly proportional to the metal - nonmetal internuclear distance in the studied minerals and rocks. The program named Agemarker can be used for the computing with the online version being available for free calculations at http://www.skyproject.org/Agemarker/Program.

To calculate $I_{av}$ index, the contents of oxides of chemical elements (in nominal mass %) are recalculated in the contents of chemical elements, being then divided into the atomic weights of the respective elements. Each quotient is multiplied by an integer - a multiplier with value being taken from 3 to 30 mln depending on the number of chemical elements in minerals and rocks and the type of the problem considered. To solve a practical problem, multiplier of 3 mln can be taken for a theoretically pure mineral, while for rocks it is recommended multiplier of 10 mln and in theoretical calculations it should be increased to 30 mln.

We obtain the numbers of atomic weights (atoms) of each element in proportion to their content in the initial sample by rounding each multiplication result to integer. For further calculations the amount of atoms of all chemical elements must be a multiple of 8, so the numbers of atoms are multiplied by 8. From this array 8 atomic weights are randomly taken, being then moved into a matrix with three rows and three columns. The ninth element of the matrix is obtained by summing the eight selected atomic weights. The information coefficient of proportionality $I_p$ is calculated for all the elements. For the rest of the array the information coefficients of proportionality are calculated similarly.

For example, we define Iav indicator for theoretically pure quartz. The initial data is entered into **Oxides Table**, or alternatively, instead of it there can be entered the contents of Si and O into **Elements table**. We set **Multiplier** of 3000000 and use default **Logarithmic base:** e (natural logarithms).

**Oxides Table**

| [№] | [Oxide] | [Content, mass %] |
|---|---|---|
| 1 | SiO2 | 100 |
| 2 | TiO2 | 0 |
| 3 | Al2O3 | 0 |
| ……………………………………… | | |
| 53 | Rb2O | 0 |

**Elements table**

| [№] | [Element] | [Atomic weight] | [Content, mass %] |
|---|---|---|---|
| 1 | H | 1.008 | 0 |
| 2 | He | 4.002602 | 0 |
| 3 | Li | 6.94 | 0 |
| ……………………………………………….. | | | |
| 95 | Am | 243.06 | 0 |

**Multiplier:** 3000000

**Logarithmic base:** e

AgeMarker calculates theoretical contents of oxygen and silicon for quartz, 53.25633% and 46.74367% respectively. After dividing these quantities by atomic weights of the corresponding elements we obtain 3.3287287 and 1.6643643. Then multiplying by 3000000 and rounding the results to integers we obtain the intermediate numbers of atomic weights 9986186 and 4993093. After multiplying these numbers by 8 we get 79,889,488 and 39,944,744 of atomic weights of oxygen and silicon respectively.

**Atomic weights (total)**

| [№] | [Atomic weight] |
|---|---|
| 1 | 0 |
| 2 | 0 |
| 3 | 0 |
| ……………………………………. | |
| 8 | 79889488 |
| ……………………………………. | |
| 14 | 39944744 |
| …………………………………… | |
| 95 | 0 |

**Atomic weights (total sum):** 119834232

The program randomly selects 8 atomic weights from the **total sum** of atomic weights and places them in a 3 x 3 matrix to determine Ip index and **Ip Squareroot** as well as frequency of the calculated Ip index. The resultant distribution is symmetrized by square root extraction. The initial values of $I_p$ indexes can be used to evaluate the effectiveness of data symmetrization. The resultant indicators are considered to be $I_{av}$ **Ip Squareroot (Average)** and **Standart deviation** (**Ip squareroot**) of **Ip Squareroot** distribution. Expected values $I_{av}$ of symmetrized distributions of sets of information coefficients of proportionality $I_p$ for atomic weights of minerals and chemical compounds are proposed to be used as main index of the calculations.

| [№] | [Ip] | [Ip Squareroot] | [Frequency] |
|---|---|---|---|
| 1 | 0.099772264 | 0.3158674786 | 584411 |
| 2 | 0.0923207361 | 0.3038432756 | 292803 |
| 3 | 0.1655827922 | 0.4069186554 | 584556 |
| 4 | 0.0905180565 | 0.3008621885 | 1168087 |
| 5 | 0.1235748975 | 0.351532214 | 1170030 |
| 6 | 0.0735236189 | 0.2711523906 | 584692 |
| 7 | 0.126016232 | 0.3549876505 | 584412 |
| 8 | 0.173626176 | 0.4166847441 | 291972 |
| 9 | 0.1420545833 | 0.3769012912 | 1168535 |
| 10 | 0.1895919516 | 0.4354215792 | 291463 |
| 11 | 0.1155315136 | 0.3398992698 | 1169486 |
| 12 | 0.1171954151 | 0.342338159 | 146268 |
| 13 | 0.1484078718 | 0.3852374226 | 876494 |
| 14 | 0.0564526157 | 0.2375975919 | 292608 |
| 15 | 0.1042890843 | 0.3229382051 | 219534 |
| 16 | 0.1101189103 | 0.3318416947 | 587580 |
| 17 | 0.0988927211 | 0.3144721308 | 146367 |
| 18 | 0.1606005409 | 0.400749973 | 291746 |
| 19 | 0.0976366955 | 0.3124687112 | 583701 |
| 20 | 0.1663667687 | 0.4078808266 | 291774 |
| 21 | 0.1796040937 | 0.4237972318 | 73175 |
| 22 | 0.1072237919 | 0.327450442 | 293292 |
| 23 | 0.1183699904 | 0.3440494011 | 146430 |
| 24 | 0.1384144667 | 0.3720409476 | 145526 |
| 25 | 0.1409563439 | 0.3754415319 | 582217 |
| 26 | 0.1315333699 | 0.3626752955 | 18409 |
| 27 | 0.0837856815 | 0.2894575643 | 291936 |
| 28 | 0.1199959733 | 0.3464043494 | 72999 |
| 29 | 0.071076287 | 0.2666013634 | 292252 |
| 30 | 0.113309201 | 0.3366143209 | 291982 |
| 31 | 0.0768448784 | 0.2772090879 | 291782 |
| 32 | 0.13476582 | 0.3671046444 | 293129 |
| 33 | 0.0923971236 | 0.3039689518 | 73008 |
| 34 | 0.1454787113 | 0.3814167161 | 73384 |
| 35 | 0.1551215649 | 0.3938547511 | 36785 |
| 36 | 0.1241362182 | 0.352329701 | 146154 |
| 37 | 0.1346859848 | 0.3669958921 | 73320 |
| 38 | 0.1149248861 | 0.3390057317 | 72852 |
| 39 | 0.041555131 | 0.2038507567 | 36952 |
| 40 | 0.1400823481 | 0.3742757647 | 146025 |
| 41 | 0.0630994573 | 0.2511960535 | 73188 |
| 42 | 0.0848703817 | 0.2913252163 | 18030 |
| 43 | 0.2067173055 | 0.4546617484 | 36361 |
| 44 | 0.0797992945 | 0.282487689 | 36574 |
| 45 | 0.1536317302 | 0.3919588374 | 18608 |
| 46 | 0.0970657487 | 0.3115537654 | 18390 |

**Ip (Average):** 0.1196745928

**Variance:** 0.0009143832

**Standart deviation:** 0.0302387703

**Ip Squareroot (Average):** 0.3430699931

**Variance (Ip squareroot):** 0.0019775727

**Standart deviation (Ip squareroot):** 0.0444699084

Minerals and rocks are characterized by unimodal distributions of $I_p$ indices, which are satisfactorily symmetrized by the extraction of square root from each coefficient. Expected values_$I_{av}$ of the symmetrized $I_p$ distributions for minerals and rocks **Ip squareroot (average)** are proposed to be used both as an indicators of the change of the potential energy of minerals and rocks, and as classification indices.

To characterize theoretically pure quartz, dolomite and calcite there were carried out $I_{av}$ indices calculations (Table 1), which were compared with the Si-O internuclear distance in α-quartz, Ca-O and MgO internuclear distances in dolomite and Ca-O one in calcite as well as with their lattices energy.

**Table 1:** Energy characteristics and Iav indices for theoretically pure quartz, calcite and dolomite

| Characteristics | Purity Silica | Purity dolomite | Purity calcite |
|---|---|---|---|
| Crystal lattice energy (kJ/moI) | 12535 [9] | 6600.8 [10] | 3163.5 [10] |
| Internuclear distance, nm | 0.161[11] | 0.2268* | 0.2381[12] |
| $I_{av}$ | 0.3430 | 0.3501 | 0.3587 |

Note. * Internuclear distance for dolomite is taken as a weighted average of Ca - O and Mg - O internuclear distances, equal to 0.2381 and 0.2081, respectively [13] on the theoretical content of CaO and MgO in dolomite.

All calculated Iav inices are directly proportional to Si-O internuclear distance in α-quartz, Ca-O and MgO internuclear distances in dolomite and Ca-O one in calcite and inversely proportional to the lattice energy. The regularities are shown grafically in Figure 1.

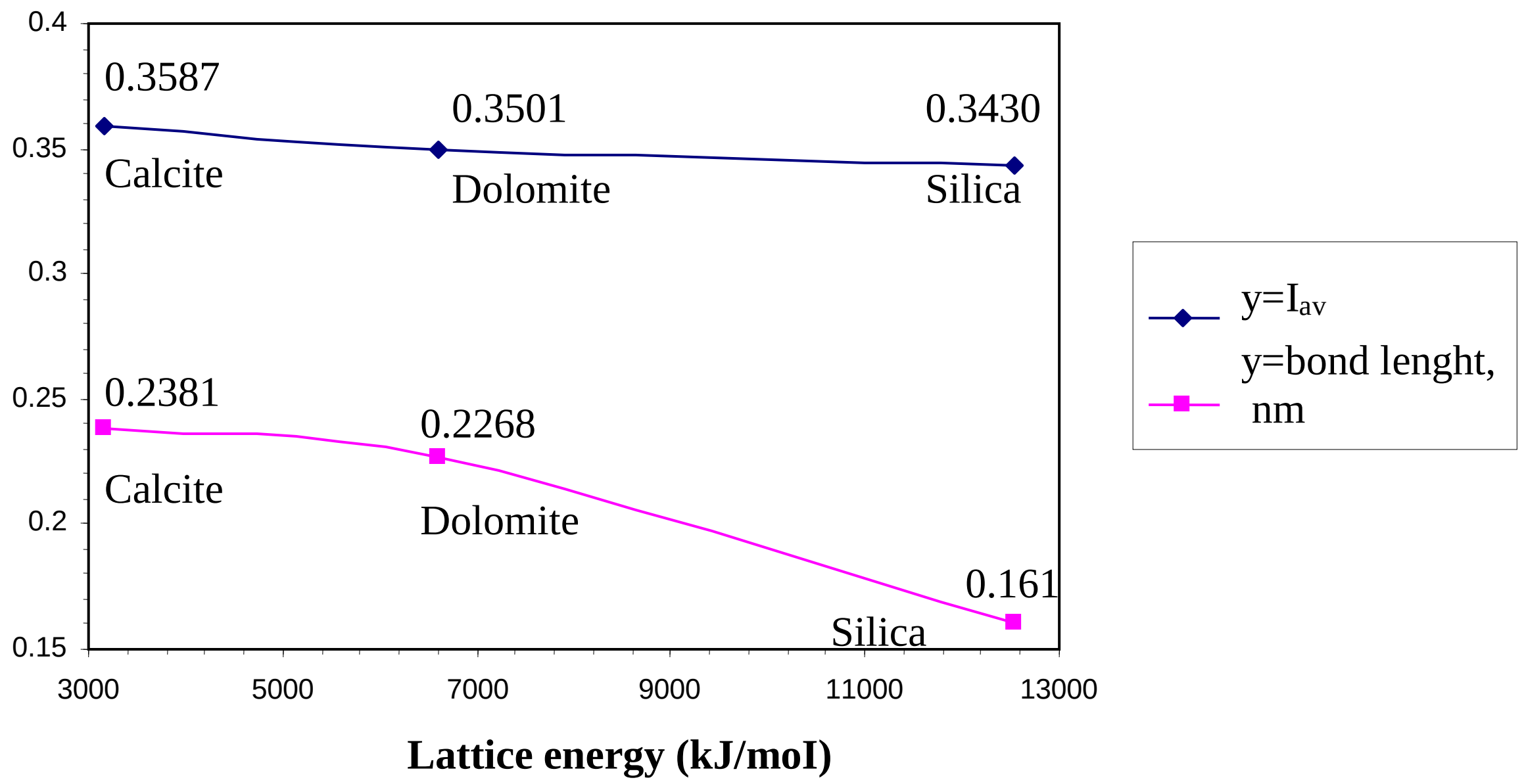

**Figure 1:** Relation between the lattice energy and indices: $\mathbf{I_{av}}$ and the **internuclear distance in molecules**.

The similar calculations were made for the reference materials of natural quartz, dolomite and limestone of the British Chemical Standard-Certified Reference Material (BCS-CRM) [14] (Table 2).

**Table 2**: Reference materials of natural quartz, dolomite and limestone BCS-CRM

| Components | Description | | | | |
|---|---|---|---|---|---|
| | High Purity Silica 313/1 (... ) | Dolomite 512 (... ) | Dolomite …(782-1) | Limestone 513 (... ) | Limestone 393 (752-1) |
| $SiO_2$ | **99.78** | **0.379** | **0.266** | **0.228** | **0.7** |
| $Al_2O_3$ | **0.036** | **0.055** | **0.104** | **0.108** | **0.12** |
| $TiO_2$ | **0.017** | **0.002** | **0.0042** | 0.004 | **0.009** |
| $Fe_2O_3$ | **0.012** | **0.03** | **0.45** | **0.0275** | **0.045** |
| MnO | **0.00013** | **0.0036** | **0.081** | **0.0095** | **0.01** |
| CaO | **0.006** | **30.61** | **30.34** | **55.59** | **55.4** |
| MgO | **0.0013** | **21.59** | **21.29** | **0.182** | **0.15** |

| Components | Description | | | | |
|---|---|---|---|---|---|
| | High Purity Silica 313/1 (... ) | Dolomite 512 (... ) | Dolomite …(782-1) | Limestone 513 (... ) | Limestone 393 (752-1) |
| $Na_2O$ | **0.003** | 0.1 | ... | <0.3 | 0.02 |
| $K_2O$ | **0.005** | <0.02 | **0.026** | **0.015** | **0.02** |
| BaO | ... | <0.02 | 0.0008 | 0.01 | **0.006** |
| $Cr_2O_3$ | <0.0002 | <0.001 | **0.0009** | **0.0012** | ... |
| PbO | ... | <0.001 | **0.0029** | **0.0009** | ... |
| ZnO | ... | <0.01 | **0.0082** | **0.0014** | ... |
| $P_2O_5$ | ... | <0.02 | **0.0128** | 0.005 | 0.005 |
| $ZrO_2$ | 0.002 | ... | ... | ... | ... |
| S | ... | <0.05 | 0.016 | **0.0097** | **0.007** |
| L.O.I. | 0.1 | **46.8** | **47.25** | **43.61** | **43.4** |
| $B_2O_3$ | ... | ... | 0.0039 | ... | ... |
| $Li_2O$ | 0.0005 | ... | ... | ... | ... |
| SrO | ... | **0.024** | ... | **0.0176** | **0.019** |
| As | ... | <0.003 | ... | <0.001 | ... |
| C | ... | 12.4 | ... | 11.9 | ... |
| Cd | ... | <0.0003 | ... | <0.001 | ... |
| F | ... | 0.01 | ... | 0.002 | <0.01 |
| Ni | ... | <0.001 | 0.0004 | <0.001 | ... |

Note. Chemical composition is given in nominal mass content in %. Figures in bold correspond to certified contents, while the rest ones characterize approximate contents. L.O.I. are set as losses on ignition and associated with the carbon content in reference materials.

For these reference materials $I_{av}$ indices are also directly proportional to the Si-O internuclear distance in α-quartz, Ca-O and MgO internuclear distance in dolomite and Ca-O one in calcite (Table 3), but natural formations differ from the theoretical pure ones in bigger $I_{av}$ indices, as their impurities reduced the lattice energy.

**Table 3:** $I_{av}$ indices for BCS-CRM reference materials of natural quartz, dolomite and limestone

| Statistics | Description | | | | |
|---|---|---|---|---|---|
| | High Purity Silica 313/1 (... ) | Dolomite 512 (... ) | Dolomite …(782-1) | Limestone 513 (... ) | Limestone 393 (752-1) |
| Iav | 0.3431 | 0.3528 | 0.3549 | 0.3612 | 0.3640 |
| Standart deviation | 0.0445 | 0.0685 | 0.0711 | 0.0816 | 0.0839 |

**RELATIVE DATING RESULTS**

We studied the records of geological sections of Cambrian, Ordovician and Upper Silurian and Lower Devonian limestones and dolomites in Virginia. [15] The results of chemical analyzes of these rocks were processed by AgeMarker program (Table 4). We analyzed the change in $I_{av}$ indices for the rocks of different ages with the same composition and for the rocks of the same age with different composition.

**Table 4:** Characteristics of Virginia limestone and dolomite

| Geologic Section No | Units No | AgeMarker Multiplier | Number of $I_p$ calculations | Iav | Standart deviation Ip | Rock | Thick-ness, feet | Mass content $SiO_2$ in % |
|---|---|---|---|---|---|---|---|---|
| 2 | 7 | 10000000 | 49904153 | 0.3586(3) | 0.0741 | Ordovician limestone | 144 | 0.33 |
| 2 | 3 and 4 | 10000000 | 49784803 | 0.3585(3) | 0.0739 | Ordovician limestone | 92 | 0.74 |
| 2 | 1 | 10000000 | 53133141 | 0.3508(8) | 0.0625 | Ordovician dolomite | 307 | 7.28 |
| 8 | 3 | 10000000 | 50025679 | 0.3584(5) | 0.0738 | Ordovician limestone | 32 | 1.28 |
| 8 | 2 | 10000000 | 50063135 | 0.3583(5) | 0.0737 | Ordovician limestone | 65 | 1.46 |
| 9 | 10 | 10000000 | 50086178 | 0.3585(0) | 0.0739 | Ordovician limestone | 70 | 0.84 |
| 9 | 9 | 10000000 | 52291947 | 0.3578(6) | 0.0729 | Ordovician limestone | 7 | 5.04 |
| 9 | 8 | 10000000 | 50012161 | 0.3584(0) | 0.0737 | Ordovician limestone | 85 | 1.32 |
| 9 | 5 | 10000000 | 49824939 | 0.3583(1) | 0.0736 | Ordovician limestone | 48 | 1.52 |
| 11 | from 1 to 3 | 10000000 | 50017304 | 0.3582(9) | 0.0736 | Ordovician limestone | 31 | 2.00 |

| Geologic Section No | Units No | AgeMarker Multiplier | Number of $I_p$ calculations | Iav | Standart deviation Ip | Rock | Thick-ness, feet | Mass content $SiO_2$ in % |
|---|---|---|---|---|---|---|---|---|
| 15 | 16 | 10000000 | 49691119 | 0.3577(0) | 0.0727 | Ordovician limestone | 109 | 6.68 |
| 15 | from 13 to 15 | 10000000 | 49847621 | 0.3583(4) | 0.0737 | Ordovician limestone | 131 | 1.52 |
| 15 | 12 | 10000000 | 49865257 | 0.3578(9) | 0.0729 | Ordovician limestone | 37 | 4.52 |
| 17 | 43, 45 and 46 | 10000000 | 49848900 | 0.3584(0) | 0.0737 | Ordovician limestone | 22 | 1.52 |
| 17 | from 1 to 29 | 10000000 | 49959146 | 0.3573(1) | 0.0721 | Ordovician limestone | 120 | 6.20 |
| 21 | 5 | 10000000 | 49892556 | 0.3584(1) | 0.0737 | Ordovician limestone | 60 | 1.12 |
| 21 | 1 | 10000000 | 50074539 | 0.3571(1) | 0.0718 | Ordovician limestone | 49 | 6.56 |
| 29 | 12 | 10000000 | 50106520 | 0.3578(0) | 0.0728 | Ordovician limestone | 40 | 4.52 |
| 29 | from 9 to 11 | 10000000 | 49993043 | 0.3583(5) | 0.0736 | Ordovician limestone | 32 | 1.64 |
| 34 | 4 и 5 | 10000000 | 49975547 | 0.3585(0) | 0.0739 | Ordovician limestone | 112 | 0.43 |
| 34 | 2 and 3 | 10000000 | 49667484 | 0.3570(1) | 0.0715 | Ordovician limestone | 48 | 9.77 |
| 34 | 1 | | 49927934 | 0.3585(4) | 0.0740 | Ordovician limestone | 25 | 1.03 |
| 36 | 30 | 10000000 | 49884089 | 0.3583(5) | 0.0737 | Ordovician limestone | 104 | 1.00 |
| 36 | from 28 to 29 | 10000000 | 50029229 | 0.3577(4) | 0.0728 | Ordovician limestone | 51 | 3.40 |
| 37 | 5 | 10000000 | 49861590 | 0.3585(0) | 0.0739 | Ordovician dolomite | 20 | 0.40 |
| 38 | 5 | 10000000 | 49773156 | 0.3584(0) | 0.0738 | * US and LD limestone | 18 | 0.73 |
| 38 | 4 | 10000000 | 50007731 | 0.3577(2) | 0.0728 | US and LD limestone | 34 | 3.55 |
| 40 | 9 | 10000000 | 49631495 | 0.3580(3) | 0.0732 | US and LD limestone | 60 | 5.05 |
| 40 | from 2 to 4 | 10000000 | 49665347 | 0.3563(8) | 0.0705 | US and LD limestone | 34 | 14.64 |
| 43 | 6 | 10000000 | 49783189 | 0.3586(4) | 0.0741 | US and LD limestone | 60 | 0.23 |
| 44 | 8 | 10000000 | 49680193 | 0.3582(7) | 0.0736 | US and LD limestone | 38 | 2.83 |
| 44 | 6 | 10000000 | 49821944 | 0.3582(6) | 0.0736 | US and LD limestone | 39 | 1.30 |
| 44 | 1 | 10000000 | 49939125 | 0.3579(8) | 0.0731 | US and LD limestone | 7 | 3.57 |
| 45 | 5 | 10000000 | 49743343 | 0.3579(0) | 0.0729 | US and LD limestone | 55 | 5.04 |

| Geologic Section No | Units No | AgeMarker Multiplier | Number of $I_p$ calculations | Iav | Standart deviation Ip | Rock | Thick-ness, feet | Mass content $SiO_2$ in % |
|---|---|---|---|---|---|---|---|---|
| 45 | 1 | 10000000 | 49978109 | 0.3580(0) | 0.0731 | US and LD limestone | 50 | 3.75 |
| 46 | 16 | 10000000 | 49940850 | 0.3580(3) | 0.0732 | US and LD limestone | 91 | 4.08 |
| 46 | 2 | 10000000 | 50064257 | 0.3577(1) | 0.0727 | US and LD limestone | 50 | 4.61 |
| 48 | from 6 to 7 | 20000000 | 99697801 | 0.3583(0) | 0.0736 | US and LD limestone | 54 | 0.58 |
| 50 | 9 and 10 | 10000000 | 53974244 | 0.3505(0) | 0.0621 | Cambrian dolomite | 298 | 3.48 |
| 50 | 6 and 7 | 10000000 | 53876383 | 0.3505(3) | 0.0621 | Cambrian dolomite | 519 | 1.44 |
| 50 | from 1 to 5 | 10000000 | 53852944 | 0.3510(5) | 0.0632 | Cambrian dolomite | 391 | 1.00 |
| 51 | from 1 to 4 | 10000000 | 54178825 | 0.3503(9) | 0.0619 | Cambrian dolomite | 603 | 1.15 |
| 52 | from 7 to 9 | 10000000 | 53854054 | 0.3504(4) | 0.0619 | Cambrian dolomite | 535 | 4.84 |
| 52 | from 1 to 6 | 10000000 | 53703292 | 0.3512(2) | 0.0634 | Cambrian dolomite | 756 | 1.00 |
| 53 | 3 | 10000000 | 53699762 | 0.3502(4) | 0.0614 | Cambrian dolomite | 530 | 9.04 |
| 53 | 2 | 10000000 | 53167694 | 0.3505(6) | 0.0619 | Cambrian dolomite | 112 | 10.28 |
| 53 | 1 | 10000000 | 53709620 | 0.3502(9) | 0.0615 | Cambrian dolomite | 155 | 7.52 |
| 54 | from 2 to 4, 6 | 10000000 | 54018910 | 0.3504(8) | 0.0620 | Cambrian dolomite | 688 | 1.03 |
| 55 | 2 | 10000000 | 53709620 | 0.3502(8) | 0.0615 | Cambrian dolomite | 336 | 3.50 |
| 55 | 1 | 30000000 | 160821349 | 0.3506(8) | 0.0623 | Cambrian dolomite | 395 | 4.48 |
| 59 | from 28 to 35 | 10000000 | 53581833 | 0.3507(8) | 0.0623 | Ordovician dolomite | 317 | 7.40 |
| 59 | from 16 to 26 | 10000000 | 53470874 | 0.3502(9) | 0.0613 | Ordovician dolomite | 296 | 11.40 |
| 59 | from 4 to 15 | 10000000 | 53697244 | 0.3504(4) | 0.0618 | Ordovician dolomite | 593 | 7.84 |
| 59 | 2 and 3 | 10000000 | 50270119 | 0.3569(9) | 0.0717 | Ordovician limestone | 490 | 5.36 |
| 61 | from 28 to 32 | 10000000 | 53631727 | 0.3507(2) | 0.0624 | Ordovician dolomite | 199 | 5.72 |
| 61 | from 18 to 27 | 10000000 | 53253881 | 0.3510(5) | 0.0628 | Ordovician dolomite | 266 | 9.48 |

| Geologic Section No | Units No | AgeMarker Multiplier | Number of $I_p$ calculations | Iav | Standart deviation Ip | Rock | Thick-ness, feet | Mass content $SiO_2$ in % |
|---|---|---|---|---|---|---|---|---|
| 66 | 83 | | 50247890 | 0.3570(9) | 0.0718 | Cambrian limestone | 125 | 6.80 |
| 67 | from 55 to 58 | 10000000 | 50528595 | 0.3553(9) | 0.0691 | Cambrian limestone | 211 | 15.08 |
| 67 | 51 and 53 | 10000000 | 50814542 | 0.3545(8) | 0.0678 | Cambrian limestone | 187 | 17.16 |
| 67 | from 46 to 50 | 10000000 | 50477733 | 0.3558(9) | 0.0699 | Cambrian limestone | 700 | 12.40 |
| 67 | from 25 to 34 | 10000000 | 53578925 | 0.3504(5) | 0.0617 | Cambrian dolomite | 298 | 9.12 |
| 67 | 23 and 24 | 10000000 | 53592717 | 0.3502(7) | 0.0614 | Cambrian dolomite | 404 | 9.44 |

Note. * US and LD limestone - Upper Silurian and Lower Devonian limestone

In general case the content of silicon dioxide in the rocks is higher than the content of any other impurity and in many cases is comparable to the total content of all impurities or exceeds it. Quartz is characterized by the maximum energy of the crystal lattice in the number of minerals: quartz - dolomite - calcite as well as small Si - O internuclear distance. We assume that higher content of silicon dioxide are always accompanied by higher potential energy of the including rock.

In a simplified version limestone and dolomite are represented as a physical system that is in equilibrium with the environment and exchanges energy and silicon dioxide with it. If in the geological sections studied we consider only Ordovician ore Silurian limestone with silicon dioxide content less than 2.83%, with the rest of them being “excluded” from the sections at all, then in most cases the one with less silicon dioxide content is proved to be younger (sections 2, 8, 9, 15, 17, 21, 34). The results of relative age determination by this rule are shown in Table 5, where correct relations are marked +, and wrong ones - .

**Table 5:** The results of limestone relative age determination by comparing silicon dioxide contents

| Geologic Section No | Units No | Iav | Rock | Mass content $SiO_2$ in % | Section_units, result of comparison (+, -) | Section_units, result of comparison (+, -) | Section_units, result of comparison (+, -) | Section_units, result of comparison (+, -) | Section_units, result of comparison (+, -) |
|---|---|---|---|---|---|---|---|---|---|

| Geologic Section No | Units No | Iav | Rock | Mass content $SiO_2$ in % | Section_units, result of comparison (+, -) | Section_units, result of comparison (+, -) | Section_units, result of comparison (+, -) | Section_units, result of comparison (+, -) | Section_units, result of comparison (+, -) |
|---|---|---|---|---|---|---|---|---|---|
| 43 | 6 | 0.3586(4) | US and LD limestone | 0.23 | | | | | |
| 2 | 7 | 0.3586(3) | Ordovician limestone | 0.33 | 43_6 + | 38_5 - | 48_6_7 - | 44_6 - | 2_3_4 + |
| 34 | 1 | 0.3585(4) | Ordovician limestone | 1.03 | 43_6 + | 38_5 - | 48_6_7 + | 44_6 - | 34_4_5 + |
| 2 | 3 and 4 | 0.3585(3) | Ordovician limestone | 0.74 | 43_6 + | 38_5 - | 48_6_7 + | 44_6 - | 2_7 + |
| 37 | 5 | 0.3585(0) | Ordovician limestone | 0.40 | 43_6 + | 38_5 - | 48_6_7 - | 44_6 - | |
| 34 | 4 и 5 | 0.3585(0) | Ordovician limestone | 0.43 | 43_6 + | 38_5 - | 48_6_7 - | 44_6 - | 34_1 + |
| 9 | 10 | 0.3585(0) | Ordovician limestone | 0.84 | 43_6 + | 38_5 + | 48_6_7 + | 44_6 - | 9_8 +<br>9_ 5 + |
| 8 | 3 | 0.3584(5) | Ordovician limestone | 1.28 | 43_6 + | 38_5 + | 48_6_7 + | 44_6 - | 8_2 + |
| 21 | 5 | 0.3584(1) | Ordovician limestone | 1.12 | 43_6 + | 38_5 + | 48_6_7 + | 44_6 - | |
| 9 | 8 | 0.3584(0) | Ordovician limestone | 1.32 | 43_6 + | 38_5 + | 48_6_7 + | 44_6 + | 9_10 +<br>9_ 5 + |
| 38 | 5 | 0.3584(0) | US and LD limestone | 0.73 | | | | | |
| 17 | 43, 45 and 46 | 0.3584(0) | Ordovician limestone | 1.52 | 43_6 + | 38_5 + | 48_6_7 + | 44_6 + | |
| 8 | 2 | 0.3583(5) | Ordovician limestone | 1.46 | 43_6 + | 38_5 + | 48_6_7 + | 44_6 + | 8_3 + |
| 29 | from 9 to 11 | 0.3583(5) | Ordovician limestone | 1.64 | 43_6 + | 38_5 + | 48_6_7 + | 44_6 + | |
| 36 | 30 | 0.3583(5) | Ordovician limestone | 1.00 | 43_6 + | 38_5 + | 48_6_7 + | 44_6 - | |
| 15 | from 13 to 15 | 0.3583(4) | Ordovician limestone | 1.52 | 43_6 + | 38_5 + | 48_6_7 + | 44_6 + | |
| 9 | 5 | 0.3583(1) | Ordovician limestone | 1.52 | 43_6 + | 38_5 + | 48_6_7 + | 44_6 + | 9_8 +<br>9_10 + |
| 11 | from 1 to 3 | 0.3582(9) | Ordovician limestone | 2.00 | 43_6 + | 38_5 + | 48_6_7 + | 44_6 + | |
| 48 | from 6 to 7 | 0.3582(9) | US and LD limestone | 0.58 | | | | | |
| 44 | 6 | 0.3582(6) | US and LD limestone | 1.30 | | | | | |

Note. The table shows the relative age determined uniquely according to geological data. The last column shows the results of comparing the relative age of the same sections with duplicated results.

The relative age of rocks is identified correctly in 53 cases out of 70; hence, the content of quartz is dependent on the age of limestone. In five sections there were made 16 measurements for Ordovician and Silurian limestones with up to 2.83% silicon dioxide contents. It was found that $I_{av}$ index is reduced by 0.0001 while increasing silicon dioxide in limestone by 0.19% on average, with variance equals to 0.027. The values of the quantity increase significantly to 0.53% average, with variance being 0.029 based on 14 measurement results, when the content of silicon dioxide in one of the compared rocks exceeds 2.83% (Table 6).

**Table 6**: Variation series of silicon dioxide content differences corresponding to limestone $I_{av}$ index change by 0.0001

| | | | | | | | | | | | | | | | | |
|---|---|---|---|---|---|---|---|---|---|---|---|---|---|---|---|---|
| Content of silicon dioxide in the rock to 2.83%* | 0 | 0 | 0 | 0.01 | 0.02 | 0.09 | 0.16 | 0.18 | 0.19 | 0.20 | 0.23 | 0.24 | 0.34 | 0.41 | 0.44 | 0.48 |
| Silicon dioxide content in one of the compared rocks over 2.83%** | 0.18 | 0.40 | 0.40 | 0.42 | 0.43 | 0.48 | 0.56 | 0.58 | 0.58 | 0.60 | 0.62 | 0.62 | 0.76 | 0.86 | | |

Note. Every difference was calculated for a pair taken of the same section. * On sections 2, 8, 9, 34 and 44; ** on sections 9, 15, 17, 21, 29, 34, 36, 38, 40, 44 and 46

Both samples don't deviate from normal distribution. Two-sample F-test for variance at significance level of 0.01 gives F = 1.07 value with critical one-sided F= 3.61. We use the variances equality for two-sample t-test with equal variances for the same significance level. We get t-statistics value 5.73 with 2.47 for critical one-sided t. Consequently, the averages of the two samples differ significantly.

Besides, the samples differ in the quality of change of silicon dioxide content, as there are three cases recorded when limestone with silicon dioxide content of up to 2.83% and the contents difference from 0.22 to 1.53% have the same value of $I_{av}$ index (sections 34, 44 and 9).

There were also made 58 and 71 similar calculations for limestone samples from different sections with silicon dioxide content up to 2.83% or more. For the sample with low silicon dioxide content we obtained the average and the variance of 0.15 and 0.024 with good approximation to normal distribution, while for the "silicified limestone" the average and the variance are 0.67 and 0.43 with a significant deviation from normal distribution. These results emphasize the similarity of processes of energy and matter exchange for all limestone with low silicon dioxide content and are indicative of more complex processes in limestone with high silicon dioxide content.

Thus, based on the example of limestone, we determined the effect of more rapid increase of potential energy of rock with relatively low lattice energy compared with "high-energy" rock. This assumed effect of potential energy change can be associated with the known Mpemba effect that deals with kinetic energy of water. The Mpemba effect states that cold water can sometimes freeze slower than warm water.

Younger Ordovician or Silurian limestone is characterized by lower content of silicon dioxide and greater value of $I_{av}$ index (sections 2, 8, 9, 34, 36, 38 and 44) compared with older limestone, if its silicon dioxide content doesn't exceed 4.5 %. The regularity is broken considering limestone with high silicon dioxide content. We do not interpret more complicated regularities because of the simplicity of our model.

The determination of relative age of rocks by chemical composition was first carried out for the Yenisei Ridge igneous rocks (south-western edge of the Siberian platform). Granites are characterized by decreasing the numerical value of $I_{av}$ index [16] subject to the decreasing of their age. Tectonic stages correspond to certain intervals of $I_{av}$ values (Table 7).

**Table 7:** Absolute age and values of $I_{av}$ index for Yenisei Ridge granites

| Maximum value $I_{av}$ | Minimum value $I_{av}$ | Absolute age, Ma |
|---|---|---|
| 0.3471(1) | 0.3463(9) | 1900-1840 |
| 0.3459(2) | 0.3454(3) | 930-700 |
| 0.3454(1) | 0.3446(8) | 650-520 |

## DISCUSSION

The method of $I_{av}$ calculating is proposed to be called Proportionality of Atomic Mass method (PAM). The changes of $I_{av}$ are correlated qualitatively and quantitatively with the variation of silicon dioxide contents in limestone; the analogy with the Mpemba effect was found. Thus we suggest that the value of $I_{av}$ index is in simple linear relation with the potential energy of minerals and rocks.

The potential energy of rock depends on the energy state of the earlier formed rocks, which transmitted energy or absorbed it to varying degrees. Using the effect it seems to be possible to identify interruptions in sedimentation and to compare energy conditions in different regions on the bases of comparing rocks of the same type.

Is it possible to use the calculation of $I_{av}$ index for chemical compounds of non crystalline nature? The calculations for some heteronuclear diatomic molecules (Table 8) allow us to answer affirmatively.

**Table 8:** Bond dissociation energy, internuclear distances and $I_{av}$ values for some heteronuclear diatomic molecules

| Characteristics | LiH | NaH | KH | RbH | HF | HCl | HBr | HI |
|---|---|---|---|---|---|---|---|---|
| Bond dissociation energy, (kcal mole$^{-1}$) [17] | 56 | 47 | 43 | 39 | 135.1 | 102.2 | 86.5 | 70.5 |
| Internuclear distance, nm | 1.5953 | 1.8873 | 2.244 | 2.367 | 0.9168 0 | 1.2744 | 1.4145 | 1.6090 |
| Iav | 0.4286 | 0.4899 | 0.5061 | 0.5209 | 0.4826 | 0.5035 | 0.5200 | 0.5254 |

$I_{av}$ indices and internuclear distances of diatomic heteronuclear molecules of alkali metal hydrides and halides are related to the bond dissociation energy by linear type dependence (Figures 2,3).

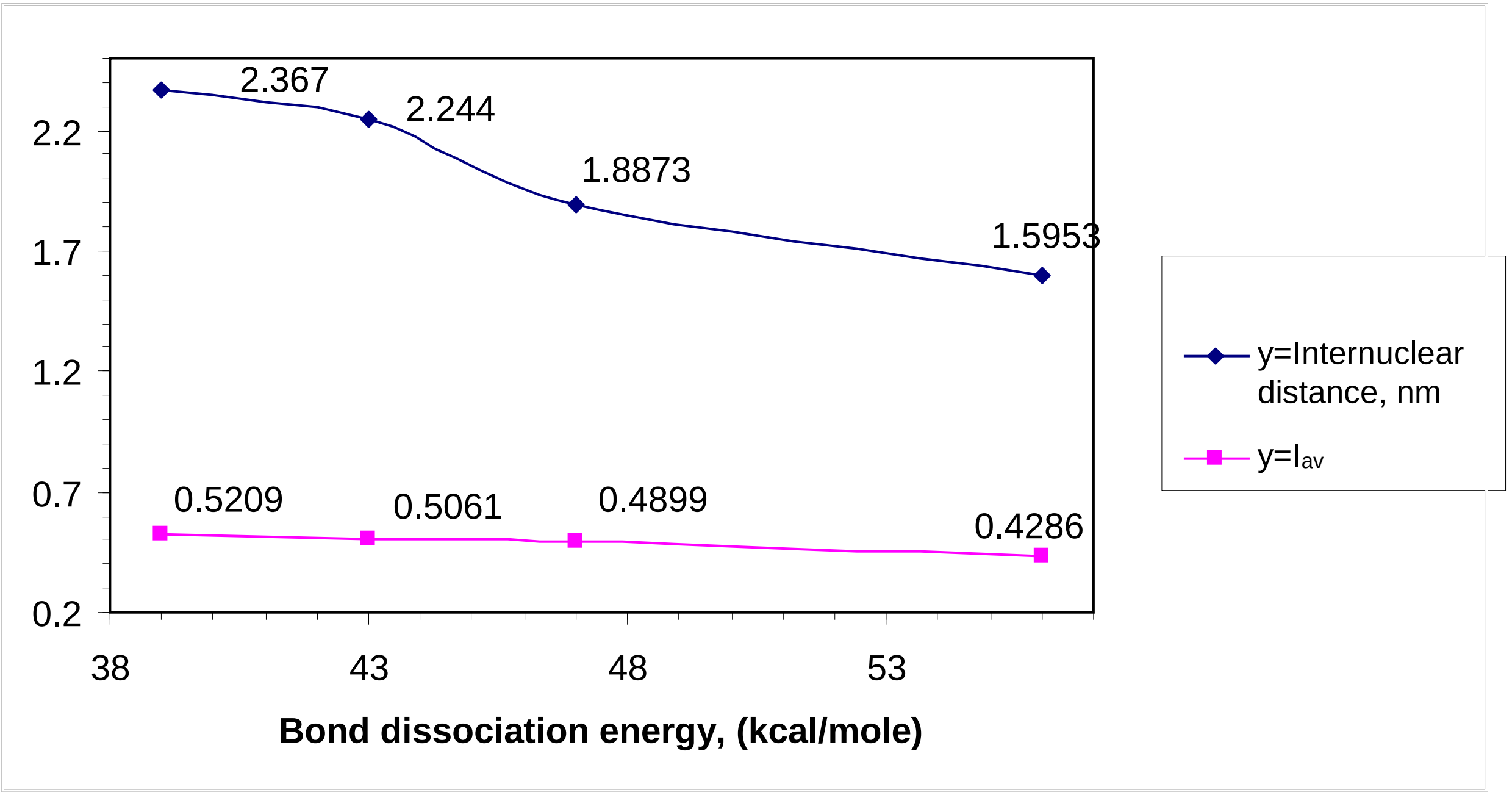

**Figure 2:** The relationship of $I_{av}$ indices and internuclear distance with potential energy in heteronuclear diatomic molecules of alkali metals hydrides

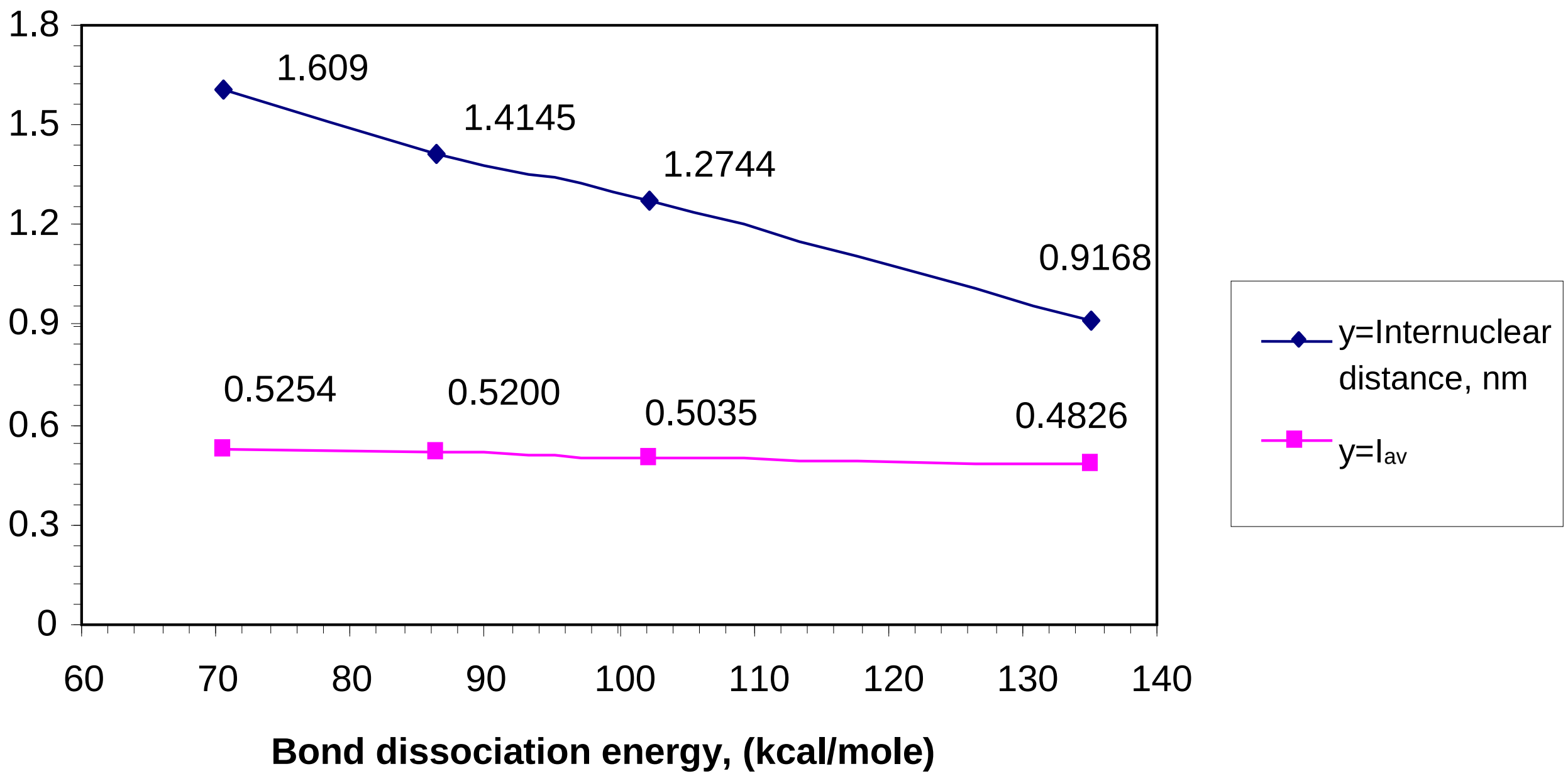

**Figure 3:** The relationship of $I_{av}$ indices and internuclear distances with potential energy in heteronuclear diatomic molecules of hydrogen halides

This relationship is similar to the one between the lattice energy and indices: «Iav» and «internuclear distances in molecules." (Figure 1).

## CONCLUSION

$I_{av}$ indices are proposed to use as indicators of minerals and rocks potential energy change related to their age, as well as new classification indicators with physical meaning. The proposed calculations give researchers important additional information about the potential energy of minerals and rocks, going beyond a laboratory. Moreover they show prospects of minerals and rocks relative dating on the basis of their chemical composition. AgeMarker can be used for theoretical research associated with the construction of the Periodic Table of Chemical Compounds [8].

## ACKNOWLEDGEMENTS

Authors thank Dr. A. Sazonov for the results of absolute dating of igneous rocks of the Yenisei Ridge. We are also grateful to Helen Phomina for a great job of proofreading the article in English and Timothy Labushev who coded AgeMarker.